\documentclass[letterpaper, 10 pt, conference]{ieeeconf}  

\IEEEoverridecommandlockouts    

\usepackage{cite}
\usepackage{amsmath,amssymb,amsfonts}
\usepackage{graphicx}
\usepackage{algorithm}
\usepackage[hidelinks]{hyperref}
\usepackage{textcomp}
\usepackage{balance}

\usepackage[utf8]{inputenc} 
\usepackage[T1]{fontenc}    
\usepackage{amsmath,amssymb,amsfonts}
\usepackage{algpseudocode}
\usepackage{todonotes}
\usepackage{multirow}
\usepackage{url}
\usepackage{cite}
\usepackage{textcomp}
\usepackage{xcolor}
\usepackage{optidef}
\usepackage{arydshln}
\usepackage{tikz}
\usepackage{circuitikz}
\usetikzlibrary{arrows.meta}
\tikzstyle{block} = [draw, rectangle, 
    minimum height=3em, minimum width=6em]
\tikzstyle{sum} = [draw, circle, node distance=1cm]
\tikzstyle{input} = [coordinate]
\tikzstyle{output} = [coordinate]
\tikzstyle{pinstyle} = [pin edge={to-,thin,black}]

\DeclareMathOperator{\Expect}{\mathbb{E}}

\newcommand{\reals}{\mathbb{R}}
\newcommand{\BBM}{\begin{bmatrix}}
\newcommand{\EBM}{\end{bmatrix}}
\newcommand{\BEQ}{\begin{equation}}
\newcommand{\EEQ}{\end{equation}}
\DeclareMathOperator{\blkdiag}{blkdiag}

\title{\LARGE \bf Maximum Likelihood Estimation of Dynamic Sub-Networks with Missing Data}
\author{João Victor Galvão da Mata$^{1}$, Anders Hansson$^{2}$, and Martin S.~Andersen$^{1}$
\thanks{*This work was supported by the Novo Nordisk Foundation under grant number NNF20OC0061894. It was also supported by ELLIIT.}
\thanks{$^{1}$Department of Applied Mathematics and Computer Science, Technical University of Denmark.
        Email: {\tt\small \{jogal,mskan\}@dtu.dk}}%
\thanks{$^{2}$Department of Electrical Engineering, Linköping University.
        Email: {\tt\small anders.g.hansson@liu.se}}%
}

\begin{document}

\maketitle
\thispagestyle{empty}
\pagestyle{empty}

\begin{abstract}

Maximum likelihood estimation is effective for identifying dynamical systems, but applying it to large networks becomes computationally prohibitive. This paper introduces a maximum likelihood estimation method that enables identification of sub-networks within complex interconnected systems without estimating the entire network. The key insight is that under specific topological conditions, a sub-network's parameters can be estimated using only local measurements: signals within the target sub-network and those in the directly connected to the so-called separator sub-network. This approach significantly reduces computational complexity while enhancing privacy by eliminating the need to share sensitive internal data across organizational boundaries. We establish theoretical conditions for network separability, derive the probability density function for the sub-network, and demonstrate the method's effectiveness through numerical examples.
 
\end{abstract}

\begin{keywords}
Dynamic networks, system identification, closed loop identification,  linear systems, statistical learning.
\end{keywords}

\section{Introduction}
\label{sec:introduction}
Dynamic networks are now prevalent in many engineering domains, including industrial systems, distributed control, reservoir engineering, and power grids \cite{dankers_thesis}. These networks present new challenges for control theory, and many existing methods for modeling, analysis, and design need to be adapted to address these challenges.

Maximum likelihood estimation (MLE) has proven to be a suitable method for identifying dynamical systems, with several works showing its advantages in comparison to other methods for different contexts, like errors-in-variables \cite{soderstromECC}, missing data \cite{wal+han14} and dynamic networks \cite{damata2023direct,HanssonEtAl2025}. However, fully identifying large networks is frequently prohibitive due to the high dimensionality of the parameter space and the burden of handling unobserved nodes. As a result, much of the literature on partial measurements focuses on identifying a single transfer function. Examples include prediction error methods (PEM) under known topology with specific measured nodes \cite{VANDENHOF20132994,Dankers_predictor_input}, Wiener-filter-based selection of sufficient signals via graph separation \cite{Materassi_unobserved_nodes,Materassi2019SignalSF}, and approaches that use auxiliary measurements to indirectly access unobserved nodes \cite{LINDER_unobservable_nodes}.

The approaches cited above often have to identify more 
transfer functions than the one that is of primary interest
in order to obtain consistent estimates. Our approach is similar
in that we identify an entire sub-network. However, we
can often obtain estimates using \emph{fewer measured signals} than traditional methods. In particular, for PEM, our numerical experiments indicate that the proposed MLE achieves comparable performance when the same signals are observed. Moreover, it remains capable of fully identifying the sub-network of interest even when fewer signals are available, a situation in which PEM cannot estimate all transfer functions.

\textbf{Contributions.} This paper introduce a maximum likelihood (ML) method for \emph{sub-network} identification in large dynamic networks:
\begin{itemize}
\item We derive conditions under which the probability density function (PDF) of the observed data depends only on the \emph{structural} parameters of a target sub-network, rendering the remaining parameters \emph{incidental} and unnecessary to estimate.
\item The formulation requires only signals within the target sub-network and within its separator sub-network, enabling substantial computational savings.
\item The same structure supports privacy-preserving estimation: independent entities can identify their respective sub-networks without sharing internal data.
\end{itemize}

Throughout the paper we assume the network topology is known. If it is not, a topology-detection step can be performed using established methods, including distance-based nonparametric approaches \cite{Materassi_topology}, Wiener-filter-based topology estimation \cite{Materassi_topology2,Dankers_topology}, causality-based detection \cite{Seneviratne_topology}, and Bayesian approaches \cite{Chiuso_topology,PVH_topology}.

\textbf{Outline.} Section~\ref{sec:dynamic-networks} introduces the network notation. Section~\ref{sec:mle} presents the separation conditions and derive the PDF for the target sub-network. Section~\ref{sec:consistency} discuss the consistency of the ML estimator, followed by numerical results and comparisons with PEM in Section~\ref{sec:num_results}. Section~\ref{sec:conclusion} concludes.

\section{Dynamic Networks}\label{sec:dynamic-networks}

Consider a network of $M$ systems where the $i$th system is described by an ARMAX model of the form
\begin{align}
\label{eqn:ARMAX}
y_k^i= -\sum_{j=1}^{n_a^i} a_j^iy_{k-j}^i  + \sum_{j=1}^{n_b^i} b_j^i u_{k-j}^i + e_k^i + \sum_{j=1}^{n_c^i} c_j^i e_{k-j}^i
\end{align}
for $k \in \{1,\ldots,N\}$. We will assume that $y_k^i$, $u_k^i$, and $e_k^i$ are zero for all $k\leq 0$. To simplify notation, we define $u^i=(u_1^i,\ldots,u_N^i)$, $y^i=(y_1^i,\ldots,y_N^i)$, and $e^i=(e_1^i,\ldots,e_N^i)$, corresponding to the $i$th system's input, output, and disturbance signals. We also define $a^i=(a_1^i,\ldots,a_{n_a^i}^i)$, $b^i=(b_1^i,\ldots,b_{n_b^i}^i)$, and $c^i=(c_1^i,\ldots,c_{n_c^i}^i)$ as well as lower-triangular Toeplitz matrices $T_{a^i}\in\reals^{N\times N}$, $T_{b^i}\in\reals^{N\times N}$ and $T_{c^i}\in\reals^{N\times N}$ whose first columns are
$$ \BBM 1\\a^i\\0\EBM,\quad\BBM 0 \\ b^i\\0\EBM,\quad\BBM 1\\c^i\\0\EBM,$$
respectively, c.f., \cite{wal+han14}. This allows us to express the ARMAX model \eqref{eqn:ARMAX} as
\begin{align}
    \label{eqn:ARMAX-T}
    T_{a^i} y^i=T_{b^i}u^i + T_{c^i} e^i.
\end{align}
The interconnections between the $M$ ARMAX systems can be defined in terms of sparse matrices $\Upsilon\in\reals^{M\times M}$ and 
$\Omega\in\reals^{M\times Q}$ with $\pm 1$ as nonzero entries such that
$$\BBM u_k^1\\u_k^2\\\vdots\\u_k^M\EBM=\Upsilon \BBM y_k^1\\y_k^2\\\vdots\\y_k^M\EBM +
\Omega \BBM r_k^1\\r_k^2\\\vdots\\r_k^Q\EBM, \quad k \in \{1,\ldots,N\},$$
where $r^j=(r_1^j,r_2^j,\ldots,r_N^j)$, $j \in \{1,\ldots,Q\}$, are exogenous signals. We will assume that $r_k^j=0$ for all $k\leq 0$.

To further simplify our notation, we define vectors $y=(y^1,\ldots,y^M)$, $u=(u^1,u^2,\ldots,u^M)$, 
$e=(e^1,e^2,\ldots,e^M)$, and $r=(r^1,r^2,\ldots,r^Q)$, and matrices
$T_{y^i}=T_{c^i}^{-1}T_{a^i}$ and $T_{u^i}=T_{c^i}^{-1}T_{b^i}$ for $i\in\{1,\ldots,M\}$. We also define two block-diagonal matrices,
\begin{align*} 
    T_y &=  \blkdiag (T_{y^1},\ldots,T_{y^M}) \\
    T_u &= \blkdiag (T_{u^1},\ldots,T_{u^M}).
\end{align*}
This allows us to express the dynamic network model as an instance of 
\begin{align}\label{non_sing_gauss_distrib_DN}
    Ax + b = \begin{bmatrix}
        e \\ 0
    \end{bmatrix},
\end{align}
with
\begin{align} 
A&=\BBM A_1\\A_2\EBM = \BBM T_y & -T_u \\ -\Upsilon\otimes I& I\EBM(P\otimes I) \label{eq:A_matrix_def}\\
x&=(P\otimes I)^T\BBM y\\u\EBM \notag\\
b&=\BBM b_1\\b_2\EBM= -\BBM 0\\ \Omega\otimes I\EBM r \notag,
\end{align}
where $A_{1}, A_{2} \in \reals^{MN\times 2MN}$, $P \in \reals^{2M \times 2M}$ is a permutation matrix that is defined such that the observed parts of $y$ and $u$ correspond to the leading entries of $x$, and $b \in \reals^{2MN}$ is partitioned conformable to $A$.

The matrix $A$ is square, and using the fact that the matrices $T_{a^i}$ and $T_{c^i}$ are non-singular for all $i$, we see that $T_y$ is non-singular. Thus, $A$ is non-singular if the Schur complement $T_y - T_u(\Upsilon\otimes I)$ is full rank. This condition is equivalent to the well-posedness of the closed loop system. 
Notice that $A_1$ depends on the unknown model parameters, 
\[\theta = (a^1,\ldots,a^M,b^1,\ldots,b^M,c^1,\ldots,c^M),\]
but $A_2$ and $b$ do not depend on $\theta$. This allows us to derive a non-singular PDF for $x$, given \eqref{non_sing_gauss_distrib_DN}, and formulate the MLE problem, as shown in \cite{damata2023direct,HanssonEtAl2025}.

\section{Network separation}\label{sec:mle}

In the following, we are not concerned with estimating the full dynamic network but rather with identifying a specific sub-network of interest inside a larger network. To formalize this setting, we introduce a graph representation that captures the interconnections among systems and enables a clear description of the sub-network separation.

\subsection{Graph Representation of Dynamic Networks}

A dynamic network of interconnected systems can be represented as a directed graph that captures the dependencies among the system outputs and external signals.

Let $\mathcal V = \{1, 2, \ldots, M+Q\}$ be the set of vertices, where vertices $1$ to $M$ correspond to the $M$ dynamical systems and vertices $M+1$ to $M+Q$ correspond to the $Q$ external signals. We consider $\mathcal V$ as the vertex set of a directed graph $\mathcal G = (\mathcal V, \mathcal E)$, with edge set $\mathcal E \subseteq \mathcal V \times \mathcal V$. The edge set is defined by the adjacency matrix $\BBM \Upsilon & \Omega \EBM^T$: there is a directed edge $(i,j) \in \mathcal E$ if and only if the output of system (or external signal)~$i$ directly affects the input of system~$j$.

Vertex $i \in \{1,\ldots,M\}$ represents the output $y^i$ of system~$i$, which depends on its own input and disturbance according to~\eqref{eqn:ARMAX-T}. The interconnection matrices $\Upsilon$ and $\Omega$ determine how outputs of other systems and external signals influence $y^i$ through the corresponding inputs $u^i$. Note that the vertices $M+1, \ldots, M+Q$, corresponding to the exogenous signal $r^1, \ldots, r^Q$, have no incoming edges and therefore act as source vertices in the graph.

We call a sequence of vertices $w = (v_1, v_2, \ldots, v_n)$ a path from vertex $v_1$ to vertex $v_n$ if $(v_i, v_{i+1}) \in \mathcal{E}$ for all $i = 1, \ldots, n-1$.

For later analysis, we partition the set of system indices into three disjoint subsets, $\{1,\ldots,M\} = A \cup B \cup C$, where $A$ denotes the \emph{target} sub-network whose parameters we wish to identify, $B$ and $C$ represents the remaining systems, where $C$ acts as a separator between them. This partition induces corresponding partitions of the signals, e.g., $y_A = \{y^i : i \in A\}$, $u_A = \{u^i : i \in A\}$, $e_A = \{e^i : i \in A\}$, and similarly for $B$ and $C$.
The external signals are partitioned according to which sub-network they excite:
\begin{align*}
    R_X = \{ r^j : (\Omega)_{ij} \neq 0 \text{ for some } i \in X \}, \quad X \in \{A,B,C\}.
\end{align*}
Note that $R_A$, $R_B$, and $R_C$ do not need to be disjoint, as a single external signal may enter multiple sub-networks. We denote by $r_A$, $r_B$, and $r_C$ the vectors formed by stacking the signals in $R_A$, $R_B$, and $R_C$, respectively.

\subsection{Separation Conditions}

We will now consider the case where we want to identify sub-network $A$ inside a larger network. The whole network is described by \eqref{non_sing_gauss_distrib_DN}, 
where $e \in \mathbb{R}^{MN}$ is a realization of a zero-mean Gaussian random variable with a diagonal covariance matrix $\Sigma_e = \blkdiag(\lambda^1 I, \dots, \lambda^M I )$, where $\lambda^i I$ is the covariance of the disturbance signal $e^i$.

We consider networks whose graph $\mathcal{G} = (\mathcal{V}, \mathcal{E})$ can be partitioned into three disjoint sub-networks, $A$, $B$, and $C$, satisfying the following structural assumptions:
\begin{enumerate}
    \item None of the outputs of systems in $B$ are connected to inputs of systems in $A$, i. e., there is no $(j,i)\in\mathcal E$ with $j\in B$ and $i\in A$.
    \item None of the outputs of systems in $A$ are connected to inputs of systems in $B$, i. e., there is no $(i,j)\in\mathcal E$ with $i\in A$ and $j\in B$.
\end{enumerate}

Under these assumptions, the vertex set $C$ separates $A$ from $B$ in $\mathcal G$, i.e., every directed path from a vertex in $A$ to a vertex in $B$ passes through $C$.

We assume that $\theta$ is ordered such that it can be partitioned as $\theta = (\theta_A,\theta_B,\theta_C)$ where $\theta_A$ are the parameters of sub-network $A$ that we wish to identify (the \textit{structural parameters}), and $\theta_B$ and $\theta_C$ are the parameters of the other sub-networks (the \textit{incidental parameters}), as in \cite{CMLE_andersen}.

We partition $x$ and $e$ conformably, such that
\begin{align*}
    x &= \BBM y_A^T & y_B^T & y_C^T & u_A^T & u_B^T & u_C^T \EBM^T,\\
    e &= \BBM e_A^T & e_B^T & e_C^T\EBM^T,
\end{align*}
with $\Sigma_e = \blkdiag(\Sigma_{eA},\Sigma_{eB},\Sigma_{eC})$.

Assumptions 1) and 2) then imply that the interconnection matrix is of the form
\begin{align*}
    \Upsilon\otimes I = \BBM \Upsilon_A & 0 & \Upsilon_{AC} \\ 0 & \Upsilon_B & \Upsilon_{BC} \\ \Upsilon_{CA} & \Upsilon_{CB} & \Upsilon_C \EBM.
\end{align*}

Defining a permutation $P$ such that 
\begin{align}\label{eq:x_bar}
    \bar x = P^Tx = \BBM y_A \\ u_A  \\ y_B \\ u_B \\ y_C \\ u_C \EBM = \BBM \bar x_A \\ \bar x_B \\ \bar x_C \EBM,
\end{align}  
and 
\begin{align*}
    P^T\BBM e \\ 0 \EBM = \BBM e_A \\ 0 \\ e_B \\ 0 \\ e_C \\ 0 \EBM,
\end{align*}  
we obtain
\begin{align}\label{eq:big_M}
 M = P^TAP = \begin{bmatrix} 
     M_{A} & 0 & M_{AC}\\
     0 & M_{B} & M_{BC}\\
     M_{CA} & M_{CB} & M_{C}
 \end{bmatrix},
\end{align}
where we use the notation
\begin{align}
   M_{X} &= \begin{bmatrix} T_{y_X}(\theta_X) & -T_{u_X}(\theta_X) \\ -\Upsilon_{X} & I\end{bmatrix}, \label{eq:M_X}\\
    M_{XY} &= \begin{bmatrix} 0 & 0 \\ -\Upsilon_{XY} & 0 \end{bmatrix}. \label{eq:M_XY}
\end{align}
We also partition $\Omega$ as
\begin{align*}
    \Omega &= \BBM \Omega_A \\ \Omega_B \\ \Omega_C \EBM,
\end{align*}
which leads to
\begin{align}
    \bar b &= P^Tb = \BBM 0 \\ -(\Omega_A\otimes I) r \\ 0 \\ -(\Omega_B\otimes I) r  \\ 0 \\ -(\Omega_C\otimes I) r \EBM = \BBM \bar b_A \\ \bar b_B \\ \bar b_C \EBM.  \label{eq:b_bar}
\end{align}

Under assumptions 1) and 2), and applying the permutation defined in \eqref{eq:x_bar}, the linear system \eqref{non_sing_gauss_distrib_DN} can be rewritten as 
\begin{align*}
    \begin{bmatrix} 
     M_{A} & 0 & M_{AC}\\
     0 & M_{B} & M_{BC}\\
     M_{CA} & M_{CB} & M_{C}
 \end{bmatrix} \BBM \bar x_A \\ \bar x_B \\ \bar x_C \EBM  + \BBM \bar b_A \\ \bar b_B \\ \bar b_C \EBM  =  \begin{bmatrix}
        e_A \\ 0 \\ e_B \\ 0 \\ e_C \\ 0
    \end{bmatrix}.
\end{align*}

From the top block row, we obtain a linear system in $\bar x_A$, 
\begin{align*}
    M_A \bar x_A + M_{AC}\bar x_C + \bar b_A = \BBM e_A \\ 0  \EBM,
\end{align*}
and using \eqref{eq:M_X}, \eqref{eq:M_XY}, \eqref{eq:b_bar}, this can be rewritten as
{\small
\begin{align}\label{eq:equivalent_net_MLE}
    \begin{split}
        \setlength{\arraycolsep}{2pt}
        \begin{bmatrix} T_{y_A}(\theta_A) & -T_{u_A}(\theta_A) \\ -\Upsilon_{A} & I\end{bmatrix} \bar x_A + \BBM 0 \\ -(\Omega_A\otimes I) r  -\Upsilon_{AC}y_C\EBM = \begin{bmatrix}
            e_A \\ 0
        \end{bmatrix}.
    \end{split}
    \end{align}
}

By treating the outputs from sub-network~$C$ as known external signals applied to sub-network~$A$ and since $e_A \sim \mathcal{N}(0,\Sigma_{eA})$, it follows that
\begin{align*}
    \setlength{\arraycolsep}{2pt}
    \bar x_A \! \sim \mathcal N\!\left(\!\! -M_A^{-1}\! \left( \!\bar b_A \!+\!\! \BBM 0 \\ -\Upsilon_{AC}y_C\EBM\!\right), M_A^{-1}\!\BBM \Sigma_{eA} & 0 \\ 0 & 0 \EBM\! M_A^{-T}\!\right).
\end{align*}
This formulation assumes that $\Upsilon_{AC}y_C$ is known and independent of $\bar x_A$. However, when feedback from sub-network~$A$ to~$C$ is present, $\Upsilon_{AC}y_C$ may depend on the same disturbances driving~$A$, and must therefore be treated as a random variable. As a result, the PDF of $\bar x_A$ obtained from the equivalent network~\eqref{eq:equivalent_net_MLE} leads to an \emph{approximate} ML formulation. In the absence of feedback from sub-network~$A$ to~$C$, this formulation leads to the \emph{true} ML.

The PDF of $\bar x_A$, when $\Upsilon_{AC}y_c$ is known, can be obtained from the equivalent network given by \eqref{eq:equivalent_net_MLE}.
Similarly, when $\Upsilon_{BC}y_c$ is known, a PDF of $\bar x_B$ can be obtained.

These results enable a valuable application of the proposed MLE: preserving data privacy during sub-network estimation. For instance, in a network representing an industrial process shared by two companies, each company can independently estimate its own sub-network while only exchanging the signals in the shared sub-network $C$, thus keeping internal data confidential.

For notational compactness, we next introduce an augmented excitation representation of the equivalent network.
Define $\hat\Omega_A$ by removing the zero columns of $\Omega_A$. There exists $\bar \Omega_A$ and $\bar \Upsilon_A$ such that $\bar \Upsilon_A \otimes I = \Upsilon_A$ and
\begin{align*}
    &\bar \Omega_A \otimes I = \BBM \hat\Omega_A \otimes I & \Upsilon_{AC} \EBM ,
    &&\bar r_A = \BBM r_A \\ y_C \EBM.
\end{align*}

Notice that, when forming the product $(\bar \Omega_A \otimes I)\bar r_A$, some components of $\Upsilon_{AC} y_C$ may be identically zero (corresponding to outputs in $C$ that do not connect to $A$). We can obtain a reduced representation by removing these zero components. Let $y_{\tilde{C}}$ denote the subvector of $y_C$ containing only the outputs that connect to $A$, and let $\tilde \Upsilon_{AC}$ be the corresponding submatrix of $\Upsilon_{AC}$ with zero rows removed, this allow us to define $\tilde \Omega_A$, such that $\tilde \Omega_A \otimes I = \BBM \hat\Omega_A \otimes I & \tilde\Upsilon_{AC} \EBM$, and
\begin{align}\label{eq:aug_r_A}
    \tilde r_A = \BBM r_A \\ y_{\tilde{C}} \EBM.
\end{align}

We can write \eqref{eq:equivalent_net_MLE} as 
\begin{align}\label{eq:equivalent_net_MLE_final}
        \setlength{\arraycolsep}{2pt}
        \begin{bmatrix} T_{y_A}(\theta_A) & -T_{u_A}(\theta_A) \\ - \bar\Upsilon_{A}\otimes I & I\end{bmatrix} \begin{bmatrix}
            y_A \\ u_A
        \end{bmatrix} + \BBM 0 \\ -(\tilde\Omega_A\otimes I) \tilde r_A\EBM = \begin{bmatrix}
            e_A \\ 0
        \end{bmatrix}.
    \end{align}

\section{Consistency of the True ML Estimator}\label{sec:consistency}

In this section, we analyze the consistency of the ML estimator obtained from the equivalent network representation in~\eqref{eq:equivalent_net_MLE_final}. The analysis is restricted to the case where the estimator corresponds to the \emph{true} ML, that is, when assumptions~1)--2) are satisfied and there is no feedback from sub-network~$A$ to sub-network~$C$. Under these conditions, $y_C$ can be treated as an exogenous signal that is independent of the disturbances in~$A$.

The equivalent network \eqref{eq:equivalent_net_MLE_final} has the same structure as the networks studied in \cite{HanssonEtAl2025}. Therefore, we can apply the MLE method from \cite{HanssonEtAl2025} to estimate the parameters $\theta_A = (a^1, \ldots, a^{|A|}, b^1, \ldots, b^{|A|}, c^1, \ldots, c^{|A|})$ of sub-network $A$, where $|A|$ denotes the number of ARMAX systems in sub-network $A$.

To establish consistency, we will make the following assumptions:
\begin{enumerate}
\item[A0.] The network is stable;
\item[A1.] There is no pole-zero cancellation in the open-loop transfer functions;
\item[A2.] The matrix $T_o \begin{bmatrix} I \\ \bar \Upsilon_A \otimes I \end{bmatrix}$ has full row rank, where $T_o$ is a selection matrix such that $T_o \bar x_A$ corresponds to the observed signals in sub-network $A$;
\item[A3.] The polynomials defined by $c^i$ in the ARMAX models have no zeros on the unit circle;
\item[A4.] $\Phi_{\tilde r_A}(\omega)\succ 0$ for almost all $\omega \in [-\pi, \pi]$, where $\Phi_{\tilde r_A}$ is the spectrum of $\tilde r_A$;
\item[A5.] Sub-network $A$ is generically identifiable, i.e., the open-loop transfer functions in $A$ can be uniquely recovered from the closed-loop transfer function from $\tilde r_A$ to the observed signals in $A$.
\end{enumerate}
\noindent\!\! Under these assumptions, \cite[Theorem 9]{HanssonEtAl2025} guarantees that the MLE estimates of the $(a,b)$ parameters, denoted by $\hat{a}_N = (\hat{a}_N^1, \ldots, \hat{a}_N^{|A|})$ and $\hat{b}_N = (\hat{b}_N^1, \ldots, \hat{b}_N^{|A|})$, based on a sample of length $N$ are consistent, i.e.,
\begin{align*}
    (\hat{a}_N, \hat{b}_N) \xrightarrow[N\to\infty]{\text{w.p. 1}} (a_0, b_0),
\end{align*}
where $(a_0, b_0)$ denotes the true $(a,b)$ parameter vectors of sub-network $A$.

\section{Numerical Example}\label{sec:num_results}

\begin{figure*}[htbp]
\vspace{0.5cm} 
\begin{center}
\resizebox{0.8\textwidth}{!}{
\begin{tikzpicture}[auto, node distance=2cm]
    \node [input] (sum11) {$+$};
    \node [block, right of=sum11, node distance=3cm] (G1) {$G^1$};
    \node [sum, right of=G1, node distance=2cm] (sum1) {$+$};
    \node [input, above of=sum1, node distance=1.2cm] (e1) {};

    \node [sum, right of=sum1, node distance=2cm] (sum12) {$+$};
    \node [block, right of=sum12, node distance=3cm] (G2) {$G^2$};
    \node [sum, right of=G2, node distance=2cm] (sum2) {$+$};
    \node [input, below of=sum2, node distance=1.2cm] (e2) {};
    \node [input, below of=sum12, node distance=1.2cm] (r2) {};
    
    \node [sum, right of=sum2, node distance=2cm] (sum13) {$+$};
    \node [input, below of=sum13, node distance=1.2cm] (r3) {};
    \node [block, right of=sum13, node distance=2cm] (G3) {$G^3$};
    \node [sum, right of=G3, node distance=2cm] (sum3) {$+$};
    \node [input, below of=sum3, node distance=1.2cm] (e3) {};
    \node [input, right of=sum3] (y3) {};


    \node [input, above of=sum3 ,node distance=1cm] (aux3) {};
    \node [input, above of=sum12, node distance=1cm] (aux4) {};


    \node [input, above of=sum3 ,node distance=2.5cm] (auxd1) {};
    \node [block, left of=auxd1, node distance=3cm] (G7) {$G^7$};
    \node [sum, left of=G7, node distance=2cm] (sum7) {$+$};
    \node [input, below of=sum7, node distance=1.2cm] (e7) {};

    \node [sum, above of=sum11, node distance=2.5cm] (sum6) {$+$};
    \node [block, right of=sum6, node distance=1.7cm] (G6) {$G^6$};
    \node [sum, right of=G6, node distance=2.5cm] (sumu6) {$+$};
    \node [input, below of=sumu6, node distance=1.2cm] (r4) {};
    \node [input, above of=sum6, node distance=1.2cm] (e6) {};


    \node [input, above of=sum7 ,node distance=1.5cm] (auxB1) {};
    \node [block, left of=auxB1, node distance=2.5cm] (G4) {$G^4$};
    \node [sum, left of=G4, node distance=1.7cm] (sum4) {$+$};
    \node [sum, above of=sumu6, node distance=1.5cm] (sum5) {$+$};
    \node [block, right of=sum5, node distance=1.7cm] (G5) {$G^5$};
    \node [input, above of=sum4, node distance=1.2cm] (e4) {};
    \node [input, above of=sum5, node distance=1.2cm] (e5) {};

    \draw [-Latex] (e1) -- node {$v^1$} (sum1);
    \draw [-Latex] (G1) -- node {} (sum1);
    \draw [-Latex] (sum1) -- node {$y^1$} (sum12);

    \draw [-Latex] (r2) -- node {$r^1$} (sum12);
    \draw [-Latex] (sum12) -- node {$u^2$} (G2);
    \draw [-Latex] (G2) -- node {} (sum2);
    \draw [-Latex] (e2) -- node {$v^2$} (sum2);
    \draw [-Latex] (sum2) -- node {$y^2$} (sum13);

    \draw [-Latex] (r3) -- node {$r^2$} (sum13);
    \draw [-Latex] (sum13) -- node {$u^3$} (G3);
    \draw [-Latex] (G3) -- node {} (sum3);
    \draw [-Latex] (e3) -- node {$v^3$} (sum3);
    \draw [-Latex] (sum3) to (aux3) to (aux4) -- node {$y^3$} (sum12);

    \draw [-Latex] (sum3) to (auxd1) -- node {$u^7 = y^3$} (G7);
    \draw [-Latex] (G7) -- node {} (sum7);
    \draw [-Latex] (e7) -- node {$v^7$} (sum7);

    \draw [-Latex] (G6) -- node {} (sum6);
    \draw [-Latex] (sumu6) -- node {$u^6$} (G6);
    \draw [-Latex] (r4) -- node {$r^3$} (sumu6);
    \draw [-Latex] (sum6) to (sum11) -- node {$u^1 = y^6$} (G1);
    \draw [-Latex] (e6) -- node {$v^6$} (sum6);
    \draw [-Latex] (sum7) -- node {$y^7$} (sumu6);


    \draw [-Latex] (sum7) to (auxB1) -- node {$u^4 = y^7$} (G4);
    \draw [-Latex] (G4) -- node {} (sum4);
    \draw [-Latex] (G5) -- node {} (sum5);
    \draw [-Latex] (e4) -- node {$v^4$} (sum4);
    \draw [-Latex] (e5) -- node {$v^5$} (sum5);
    \draw [-Latex] (sum5) -- node {$y^5$} (sumu6);
    \draw [-Latex] (sum4) -- node {$u^5 = y^4$} (G5);

\end{tikzpicture}}
\end{center}
\caption{Block diagram for the example network; We use $v^i = H^ie^i$.}
\centering
\label{fig:network_num_example}
\end{figure*}
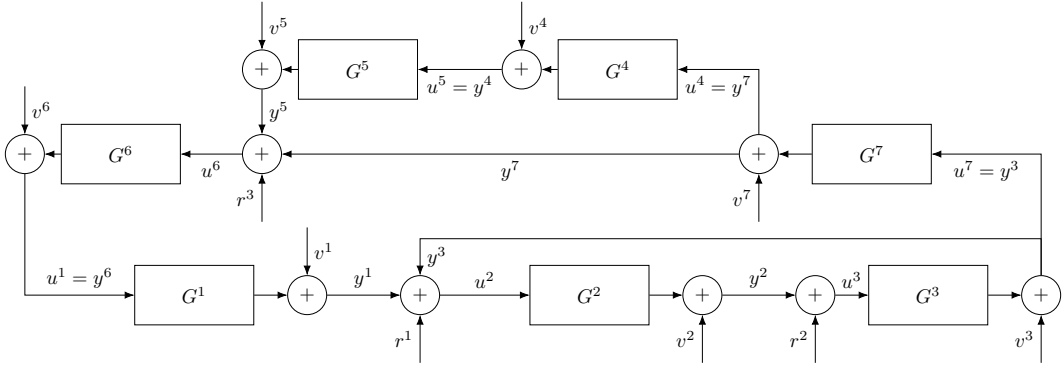

We will now demonstrate some properties of the proposed approach through a numerical example. We will use the example network shown in Figure~\ref{fig:network_num_example}, for which the interconnections may be characterized by the matrices\begin{align*}
    \Upsilon = \BBM
    0 & 0 & 0 & 0 & 0 & I & 0 \\
    I & 0 & I & 0 & 0 & 0 & 0 \\
    0 & I & 0 & 0 & 0 & 0 & 0 \\
    0 & 0 & 0 & 0 & 0 & 0 & I \\
    0 & 0 & 0 & I & 0 & 0 & 0 \\
    0 & 0 & 0 & 0 & I & 0 & I \\
    0 & 0 & I & 0 & 0 & 0 & 0 \\
    \EBM, \quad \Omega =  \BBM
    0 & 0 & 0 \\
    1 & 0 & 0 \\
    0 & 1 & 0 \\
    0 & 0 & 0 \\
    0 & 0 & 0 \\
    0 & 0 & 1 \\
    0 & 0 & 0 \\
    \EBM.
\end{align*}
For this network, we will be interested in identifying $\mathbf G^1$, $\mathbf G^2$ and $\mathbf G^3$, and then we will have $A = \{1,2,3\}$, $B = \{4,5\}$ and $C = \{6,7\}$. The equivalent sub-network $A$ is of the form \eqref{eq:equivalent_net_MLE_final} with
\begin{align*}
    \bar\Upsilon_A = \BBM 0 & 0 & 0 \\ 1 & 0 & 1 \\ 0 & 1 & 0 \EBM, \quad \tilde \Omega_A = \BBM 0 & 0 & 1 \\ 1 & 0 & 0 \\ 0 & 1 & 0 \EBM, \quad \tilde r_A = \BBM r_1 \\ r_2 \\ y_6 \EBM.
\end{align*}

The true discrete transfer functions used to generate the data are
\begin{align*}
    \mathbf G_1 &= \frac{0.3z+0.15}{z^2+z+0.25},
    &&\mathbf G_2 = \frac{0.8z-0.3}{z^2-0.8z+0.15}, \\
    \mathbf G_3 &= \frac{-0.4z-0.25}{z^2+0.45z-0.13},
    &&\mathbf G_4 = \frac{-2z+0.4}{z^2-0.45z-0.1}, \\
    \mathbf G_5 &= \frac{2.2z+2}{z^2+0.1z-0.4},
    &&\mathbf G_6 = \frac{0.15z+0.05}{z^2-0.2z-0.15}, \,\\
    \mathbf G_7 &= \frac{z+0.2}{z^2+0.5z+0.05},
\end{align*}
and 
\begin{align*}
    \mathbf H_1 &= \frac{z^2+0.3z-0.01}{z^2+z+0.25},
    &&\mathbf H_2 = \frac{z^2-0.8z+0.2}{z^2-0.8z+0.15}, \\
    \mathbf H_3 &= \frac{z^2-0.02z-0.8}{z^2+0.45z-0.13},
    &&\mathbf H_4 = \frac{z^2-0.15z-0.07}{z^2-0.45z-0.1}, \\
    \mathbf H_5 &= \frac{z^2-0.6z-0.05}{z^2+0.1z-0.4},
    &&\mathbf H_6 = \frac{z^2+z+0.15}{z^2-0.2z-0.15}, \,\\
    \mathbf H_7 &= \frac{z^2+0.15z-0.7}{z^2+0.5z+0.05}.
\end{align*}
We can check that these transfer functions yield a stable network.

To evaluate the model fit for a given parameter estimate $\hat \theta_A$, we will use the fit values for the simulated outputs of each internal transfer function in $A$. Given $\hat \theta_A$ and new realizations of the error signals, $e$, and the inputs, $r$, the fit is defined as
\begin{align*}
    \text{fit}(\hat x) = 1 - \frac{\|\hat x - x^{\mathrm{ref}}\|_2}{\|x^{\mathrm{ref}} - \mathbf{1}\bar x^{\mathrm{ref}} \|_2},
\end{align*}
where $\hat x$ is the signal of interest from the estimated model defined by $\hat \theta_A$, $x^{\mathrm{ref}}$ is the same signal from the true model, $\bar x^{\mathrm{ref}}$ denotes the mean of the true signal $x^{\mathrm{ref}}$, and $\mathbf{1}$ is a vector of all ones.

Since we just estimate sub-network $A$, i.e. we only have $\hat \theta_A$, to compute the fit values, we will simulate the network using $\hat \theta_A$ and the true value for $y^6$, which is the signal coming from the separator.

We compare the proposed MLE approach with a traditional PEM approach. We generated $N=500$ measurements where the errors $(e^1,e^2,e^3,e^4,e^5,e^6,e^7)$ are realizations of a zero-mean Gaussian random variable with covariance $\Sigma_e = \blkdiag(0.01 I, 0.02 I, 0.03 I, 0.04 I, 0.05 I, 0.06 I, 0.07 I)$, and the inputs $(r^1,r^2,r^3)$ are vectors of independent samples from the Rademacher distribution. To compute the fit values, we utilized the same validation data for both methods, which was generated in the same manner as the estimation data. The System Identification Toolbox \cite{ljung1995system} in MATLAB R2024a was used to compute the estimates using the PEM.

The optimization problem associated with the ML approach is non-convex. To reduce the risk of getting stuck in a poor local minima we use an initialization strategy. We first solve an ML problem where we consider $T_{c^i} = I$ in \eqref{eqn:ARMAX-T}, i.e. we approximate each system with an ARX model and we also constrain all $\lambda^i$ to be the same. Then, we again solve an ML problem, but this time for an ARMAX model. We still constrain all $\lambda^i$ to be the same, and we initialize the solver with the value of $(a,b)$ we obtained in the previous step. This will provide us with initial values $(a,b,c,\lambda)$ for a third  final step, where we do not constrain $\lambda^i$.  We solve all the optimization problems with a trust-region method implemented via SciPy’s optimize module. Our implementation is based on the Python library JAX \cite{jax}, which makes use of automatic differentiation to compute the partial derivatives of the cost function. Our code and data are available on GitHub\footnote{Available at: \url{https://github.com/Jvictormata/mle_sub_nets}}.

We can check that if we only observe $y^3$, together with the external signals  $(y^6,r^1,r^2)$, we have
\begin{align*}
    \mathbf y^3 &= \mathbf G^3 \mathbf G^2 \mathbf G^1\mathbf y^6 + \mathbf G^3\mathbf G^2(\mathbf y^3 + \mathbf r^1) + \mathbf G^3 \mathbf r^2 \\ &\quad + \mathbf G^3 \mathbf G^2 \mathbf H^1 \mathbf e^1 + \mathbf G^3 \mathbf H^2\mathbf e^2 + \mathbf H^3\mathbf e^3.
\end{align*}
The closed loop transfer function from $(y^6,r^1,r^2)$ to $y^3$ will be
\begin{align*}
    \mathbf G_c = \BBM \mathbf \triangle^{-1}\mathbf G^3\mathbf G^2 \mathbf G^1 & \mathbf \triangle^{-1}\mathbf G^3\mathbf G^2 & \mathbf \triangle^{-1}\mathbf G^3\EBM,
\end{align*}
where $\mathbf \triangle = (1-\mathbf G^3\mathbf G^2)$. The open loop transfer functions can then be uniquely determined from $\mathbf G_c$ from 
\begin{align*}
    \mathbf G^1 = \frac{\mathbf G_{c_1}}{\mathbf G_{c_2}}, \quad \mathbf G^2 = \frac{\mathbf G_{c_2}}{\mathbf G_{c_3}}, \quad \mathbf G^3 = \frac{\mathbf G_{c_3}}{1+\mathbf G_{c_2}}.
\end{align*}
This ensures generic network identifiability and allows us use the MLE approach to determine all the three open loop transfer functions. Notice that with PEM we would only be able to directly estimate $\mathbf G^3$ for this case.

For the cases where we also measure $y^1$, we have
\begin{align*}
     \mathbf y^1 &= \mathbf G^1\mathbf y^6 + \mathbf H^1\mathbf e^1\\
     \mathbf y^3 &= \mathbf G^3 \mathbf r^2 + \mathbf G^3\mathbf G^2(\mathbf y^1 + \mathbf y^3 + \mathbf r^1) + \mathbf H^3\mathbf e^3 + \mathbf G^3\mathbf H^2\mathbf e^2,
 \end{align*}
which only allows us to directly recover $\mathbf G^1$ and $\mathbf G^3$ with PEM.

We are only able to directly recover all the open loop transfer functions with PEM when observing $(y^1,y^2,y^3)$ together with the external signals  $(y^6,r^1,r^2)$. Leading to the equations
\begin{align*}
     \mathbf y^1 &= \mathbf G^1\mathbf y^6 + \mathbf H^1\mathbf e^1\\
     \mathbf y^2 &= \mathbf G^2(\mathbf r^1 + \mathbf y^1 + \mathbf y^3) + \mathbf H^2\mathbf e^2\\
     \mathbf y^3 &= \mathbf G^3 (\mathbf r^2 + \mathbf y^2) + \mathbf H^3\mathbf e^3.
 \end{align*}

For both methods, the covariance,
\begin{align*}
    \text{Cov}(\hat \theta ) &= \Expect(\hat \theta -  \Expect \hat\theta)(\hat \theta -  \Expect\hat\theta)^T,
\end{align*}
and the bias,
\begin{align*}
    \text{bias}(\hat \theta ) =  \Expect \hat\theta - \theta_0,
\end{align*}
were computed numerically using averages, by performing 100 estimates based on different realizations of $e$ with the same values for $r$. We will compute both the fit and the covariance only for the $(a,b)$ parameters.

\begin{table*}[htbp]
    \centering
  \hspace*{\dimexpr -\oddsidemargin}
    \caption{Comparison of the PEM and MLE: fit values and trace and max. eigenvalue of the covariance matrix.}
    \label{tab:table_PEM_vs_MLE}
   \setlength\tabcolsep{3pt}
   \scalebox{0.95}{
  \begin{tabular}{|c|c|c|c|c|c|c|c|c|c|c|c|c|c|}
    \hline
    \multirow{3}{*}{\shortstack{Observed\\ signals}} &
      \multicolumn{6}{|c|}{Direct method (PEM)} &
      \multicolumn{6}{|c|}{Proposed MLE method} \\
      & \multicolumn{3}{|c|}{$100\times$} & \multicolumn{2}{|c|}{Covariance matrix} & \multirow{2}{*}{$\|\text{bias}\|_2$} & \multicolumn{3}{|c|}{$100\times$} & \multicolumn{2}{|c|}{Covariance matrix} & \multirow{2}{*}{$\|\text{bias}\|_2$}\\
    & $\text{Fit}(\hat y^1)$ & $\text{Fit}(\hat y^2)$ & $\text{Fit}(\hat y^3)$ & Trace & Max.\ eigenvalue & & $\text{Fit}(\hat y^1)$ & $\text{Fit}(\hat y^2)$ & $\text{Fit}(\hat y^3)$ & Trace & Max. eigenvalue &  \\
    \hline
     $y^3$ & -- & -- & -- & -- & --  & -- & 56.58 & 75.89 & 60.36  & 0.3764 & 0.1516 & 1.3776 \\
    \hline
     $y^1$ and $y^3$ & -- & -- & -- & -- & -- & -- & 57.10 & 75.84 & 60.21 & 0.2442 & 0.1548 & 1.2648 \\
    \hline
    $y^1, y^2$ and $y^3$ & 57.13 & 76.86 & 60.43 & 0.8717 & 0.4894 & 0.1400 & 57.13 & 76.86 & 60.43 & 0.8501 & 0.5148 & 0.1974 \\
    \hline
  \end{tabular}}
\end{table*}

Table~\ref{tab:table_PEM_vs_MLE} summarizes the results. With PEM, all transfer functions can be identified only when the measurements include $(y^1,y^2,y^3)$ together with the external signals $(y^6,r^1,r^2)$. For cases with fewer measured signals, we only show the results for the MLE approach.

We can check that when observing the same signals, the MLE method yields similar results as PEM. The main advantage of the proposed MLE approach is that it can fully identify the network based on fewer observed signals, with only a small decrease in the fit values.

In this numerical example, we observed a larger bias when fewer signals are measured; this was mainly due to the optimization procedure rather than the estimation method itself. Because the ML cost function is nonconvex, our initialization strategy caused some of the 100 realizations to converge to local minima far from the true parameters. When using more informative initializations (e.g., starting near the true parameters or applying multiple restarts), the bias was significantly reduced, confirming that part of it originates from optimization rather than model misspecification. Moreover, since the network in this example includes feedback from sub-network $A$ to $C$, the estimator corresponds to the \emph{approximate} ML formulation, for which consistency has not been established, and hence some bias could be present.

\section{Conclusion}\label{sec:conclusion}

This paper has introduced an MLE framework for identifying sub-networks within large dynamic networks. By exploiting PDFs that depend only on the parameters of the sub-network of interest, MLE enables estimation without involving the incidental parameters of the rest of the network, which leads to  significant computational savings by avoiding the estimation of large-scale models.

In our numerical experiments, MLE achieved results comparable to PEM whenever both methods apply, while being able to fully identify the sub-network of interest even in cases where PEM is not able to estimate all transfer functions of the internal systems.

In this work, we established consistency results only for the case where the equivalent network~\eqref{eq:equivalent_net_MLE} yields the true ML estimator. Extending these results to the approximate ML setting remains an important direction for future research.
Also, relating our approach to other consistent 
methods for identifying a single 
transfer functions or a sub-network of 
transfer functions is an interesting future topic.

\balance 

\bibliographystyle{plain}        

\end{document}